\documentclass[runningheads]{llncs}
\usepackage[year=2025,ID=*****]{iciap}
\usepackage{iciapabbrv}
\usepackage{graphicx}
\usepackage{booktabs}
\usepackage[accsupp]{axessibility}  % Improves 
\usepackage[pagebackref,breaklinks,colorlinks,citecolor=iciapblue]{hyperref}
\usepackage{orcidlink}
\begin{document}

\title{Overfitting in Histopathology Model Training: The Need for Customized Architectures} 
\titlerunning{Overfitting in Histopathology}

\author{Saghir Alfasly \and Ghazal Alabtah \and H.R. Tizhoosh\orcidlink{0000-0001-5488-601X}}

\authorrunning{S.~Alfasly et al.}
\institute{KIMIA Lab, Department of Artificial Intelligence \& Informatics, Mayo Clinic, 
Rochester, MN, USA}

\maketitle

\begin{abstract}
This study investigates the critical problem of overfitting in deep learning models applied to histopathology image analysis. We show that simply adopting and fine-tuning large-scale models designed for natural image analysis often leads to suboptimal performance and significant overfitting when applied to histopathology tasks. Through extensive experiments with various model architectures, including ResNet variants and Vision Transformers (ViT), we show that increasing model capacity does not necessarily improve performance on histopathology datasets. Our findings emphasize the need for customized architectures specifically designed for histopathology image analysis, particularly when working with limited datasets. Using Oesophageal Adenocarcinomas public dataset, we demonstrate that simpler, domain-specific architectures can achieve comparable or better performance while minimizing overfitting.
\end{abstract}

\section{Introduction}
Recent advances in artificial intelligence, particularly deep learning, have shown tremendous potential in various domains, including medical image analysis. Histopathology, the microscopic examination of tissue samples, is a critical field in medical diagnostics, particularly digital pathology, which will greatly benefit from these technological advances. However, the unique characteristics of histopathology images and their complex tissue patterns present specific challenges that are not adequately addressed by models designed for non-medical imaging, i.e., natural image analysis \cite{morid2021scoping,xie2018pre,li2022improving}.

Current approaches in computational pathology often rely on adopting and fine-tuning existing models originally developed for natural image classification tasks \cite{riasatian2021fine,sharmay2021histotransfer}. Although this approach has yielded some success, it often lacks the ability to address the unique challenges presented by histopathology images, such as high resolution, multiscale features, and specific staining patterns \cite{sikaroudi2023generalization,chang2021stain}.

\begin{figure}[htbp]
\centering
\includegraphics[width=0.98\textwidth]{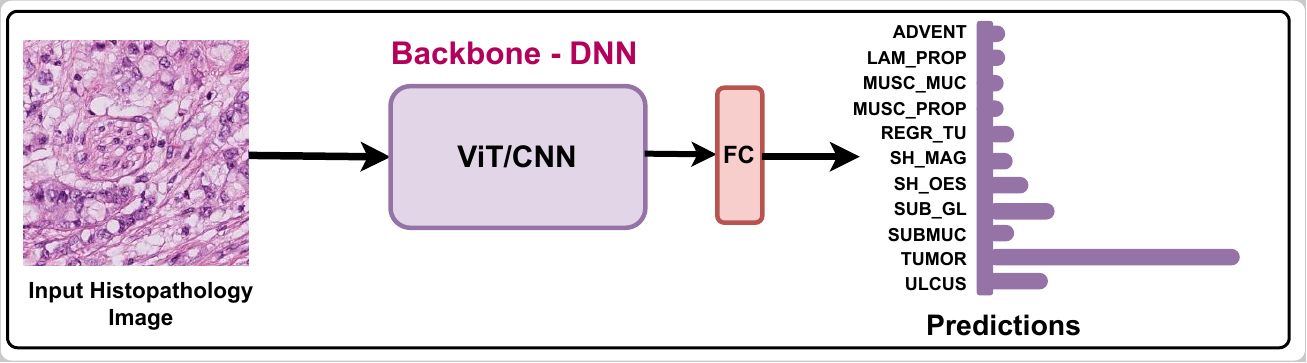}
\caption{Illustration of the classification pipeline. A histopathology image is fed to the trained model variant which has been trained in a supervised paradigm on Oesophageal Adenocarcinomas Dataset. The model provides the output classes' probability.}
\label{fig:overview}
\end{figure}

In this paper, we investigate the problem of overfitting in deep learning models \cite{li2019research, bejani2021systematic} applied to histopathology image analysis  \cite{banerji2022deep, wu2022recent, alfasly2024rotation} through a set of systematic experiments with variant deep neural networks in terms of their depth and width, as depicted in Fig. \ref{fig:overview}. We hypothesize that the capacity of models designed for natural images may be excessive for histopathology tasks, leading to overfitting and poor generalization. Through a series of experiments, we demonstrate the limitations of current models and argue for the development of customized architectures specifically designed for histopathology image analysis.

\section{Methods}
\subsection{Dataset}
\begin{figure}[htbp]
\centering
\includegraphics[width=0.98\textwidth]{ 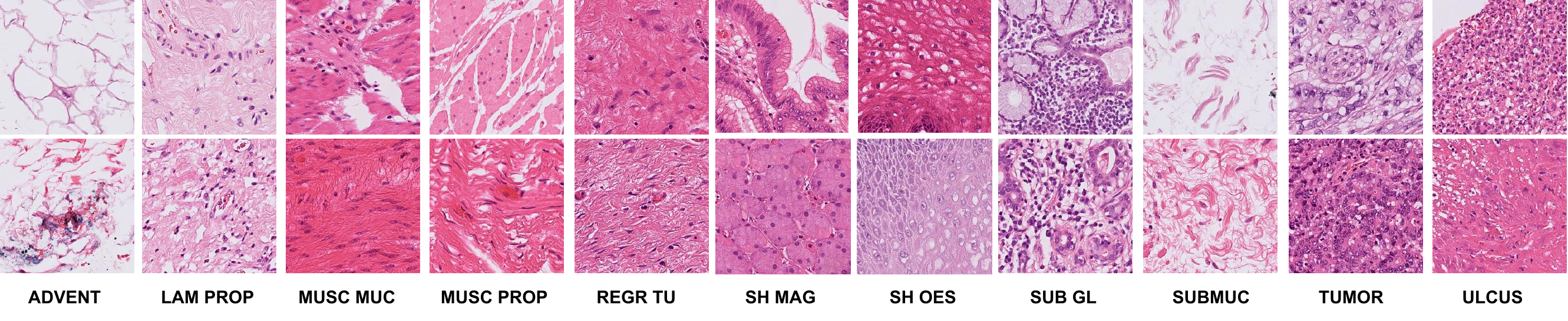}
\caption{Hisopathology image samples for each class in oesophageal adenocarcinomas dataset \cite{tolkach2023artificial}. There are ``patches'', i.e., sub-images,  extracted from gigapixel whole slide images that have extremely large dimensions.}
\label{fig:imageSamples}
\end{figure}

We used the Oesophageal Adenocarcinomas dataset, which comprises histopathological images from multiple medical centers \cite{tolkach2023artificial
}. The data set includes 367,457 total images collected from The Cancer Genome Atlas (SET3-TCGA), the University Hospital Cologne (SET4-CHA), the University Hospital Berlin-Charité (SET1-UKK) and the Landesklinikum Wiener Neustadt (SET2-WNS), as detailed in Table \ref{tab:dataset}. The images were digitized using standard histopathology scanners at a magnification of 40x and pre-processed to a uniform size of 256×256 pixels. The training set combines 178,219 images from SET4-CHA and 32,828 images from SET3-TCGA, providing diverse examples of adenocarcinoma tissue patterns. The validation set (SET1-UKK) contains 34,736 images, while the test set (SET2-WNS) includes 121,674 images, ensuring a robust evaluation of model performance. All images were annotated by experienced pathologists and include various histological patterns characteristic of oesophageal adenocarcinomas. A set of image samples for each class is shown in Fig. \ref{fig:imageSamples} (URL: https://zenodo.org/records/7548828).

\begin{table}[htbp]
\caption{The data splits of Oesophageal Adenocarcinomas Dataset.}
\label{tab:dataset}
\begin{center}
\begin{tabular}{|l|l|r|}
\hline
\textbf{Dataset} & \textbf{Phase} & \textbf{Size} \\
\hline
SET4 CHA & training & 178,219 \\
SET3 TCGA & training & 32,828 \\
SET1 UKK & validation & 34,736 \\
SET2 WNS & test & 121,674 \\
\hline
\end{tabular}
\end{center}
\end{table}

\subsection{Models' Architectures}
We conducted a systematic investigation of various deep learning architectures to understand the relationship between model capacity and performance in histopathological image analysis. Our study encompassed three major architectural families: ResNet variants, Vision Transformers, and a baseline SimpleCNN model.
\subsubsection{ResNet Variants}
We evaluated eight different ResNet \cite{he2016deep} configurations with varying depths and complexities. Starting from a minimal ResNet with [1,1,1,1] blocks, we progressively increased the network capacity to ResNet-152 with [3,8,36,3] blocks. Each configuration follows the standard ResNet architecture with residual connections, but differs in the number of blocks per stage.
The basic building block consists of two 3×3 convolutional layers with batch normalization and ReLU activation. In deeper variants (ResNet-50 and above), we used the bottleneck design with 1×1 convolutions for dimensionality reduction and expansion. Our tested configurations ranged from the basic ResNet with [1,1,1,1] configuration (4 stages with single residual blocks), through balanced architectures like ResNet-18 [2,2,2,2], to progressively deeper variants. These included modified designs with increasing depth in the middle stages [1,2,3,2], uniform depth distributions [3,3,3,3], and standard deep architectures ResNet-50 [3,4,6,3], ResNet-101 [3,4,23,3], and ResNet-152 [3,8,36,3]. Each variant maintained the standard ResNet downsampling pattern, halving spatial dimensions and doubling channels after each stage. We initialized all models with ImageNet pre-trained weights and fine-tuned them on our histopathology dataset.
\subsubsection{Vision Transformers}
Our Vision Transformer (ViT) \cite{dosovitskiy2020image} experiments explored three key architectural dimensions: network depth, embedding size, and number of attention heads. We systematically varied these parameters to understand their impact on model performance and overfitting tendencies.

For network depth investigation, we examined configurations ranging from 1 to 24 transformer layers while keeping embedding size (384) and number of heads (6) constant. Each transformer layer incorporated a multi-head self-attention mechanism, layer normalization before both attention and MLP blocks, a two-layer MLP with GELU activation, and residual connections surrounding both the attention and MLP components.

The embedding dimension experiments spanned from 24 to 1536 dimensions, maintaining a constant depth of 6 layers and 6 attention heads. Our embedding process involved converting 16×16 image patches to tokens, incorporating learnable 1D position encodings, and using a special [CLS] token for classification. The attention head variations ranged from 2 to 16 heads while keeping other parameters fixed (depth=6, embedding size=768). Each attention mechanism implemented parallel attention heads with scaled dot-product attention, using dimension = embedding-size/num-heads, along with linear projections for queries, keys, and values, and an output projection combining all head outputs.

For all ViT variants, we employ comprehensive training techniques that include dropout (rate = 0.1) in the attention and MLP layers, initialization of the layer scale for stable training, stochastic depth for regularization, and gradient clipping to prevent explosion.

\subsubsection{SimpleCNN}
As a baseline comparison, we implemented a straightforward 6-layer CNN architecture. This network served as a control to evaluate whether complex architectures provide meaningful benefits over simpler designs. The SimpleCNN architecture begins with an input layer using 3×3 convolution with 64 filters, followed by four main layers each comprising 3×3 convolution, batch normalization, ReLU activation, and 2×2 max pooling. The fifth layer maintains the same structure without pooling, and the architecture concludes with global average pooling and a fully connected output layer.

Our training methodology remained consistent across all architectures. We used the AdamW optimizer with a weight decay of 0.01 and a learning rate of 0.0001 following a cosine decay schedule. Training proceeded with a batch size of 32, incorporating standard data augmentation techniques that include random flips, rotations, and color jittering. We implemented early stopping based on validation loss with a patience of 15 epochs to prevent overfitting. This comprehensive set of architectures and configurations allowed us to systematically investigate the relationship between model capacity and performance in histopathology image analysis while maintaining consistent training procedures across all experiments.

\section{Results}

\subsection{ResNet Performance}
The performance analysis of ResNet variants revealed several interesting patterns regarding model capacity and overfitting. Table \ref{tab:resnet} presents comprehensive results across different ResNet configurations. The simplest ResNet variant with [1,1,1,1] blocks achieved a validation F1 score of 0.73, while the most complex ResNet-152 with [3,8,36,3] blocks achieved a similar validation F1 score of 0.73, despite having significantly more parameters.

\begin{figure}[htbp]
\centering
\includegraphics[width=0.5\textwidth]{ 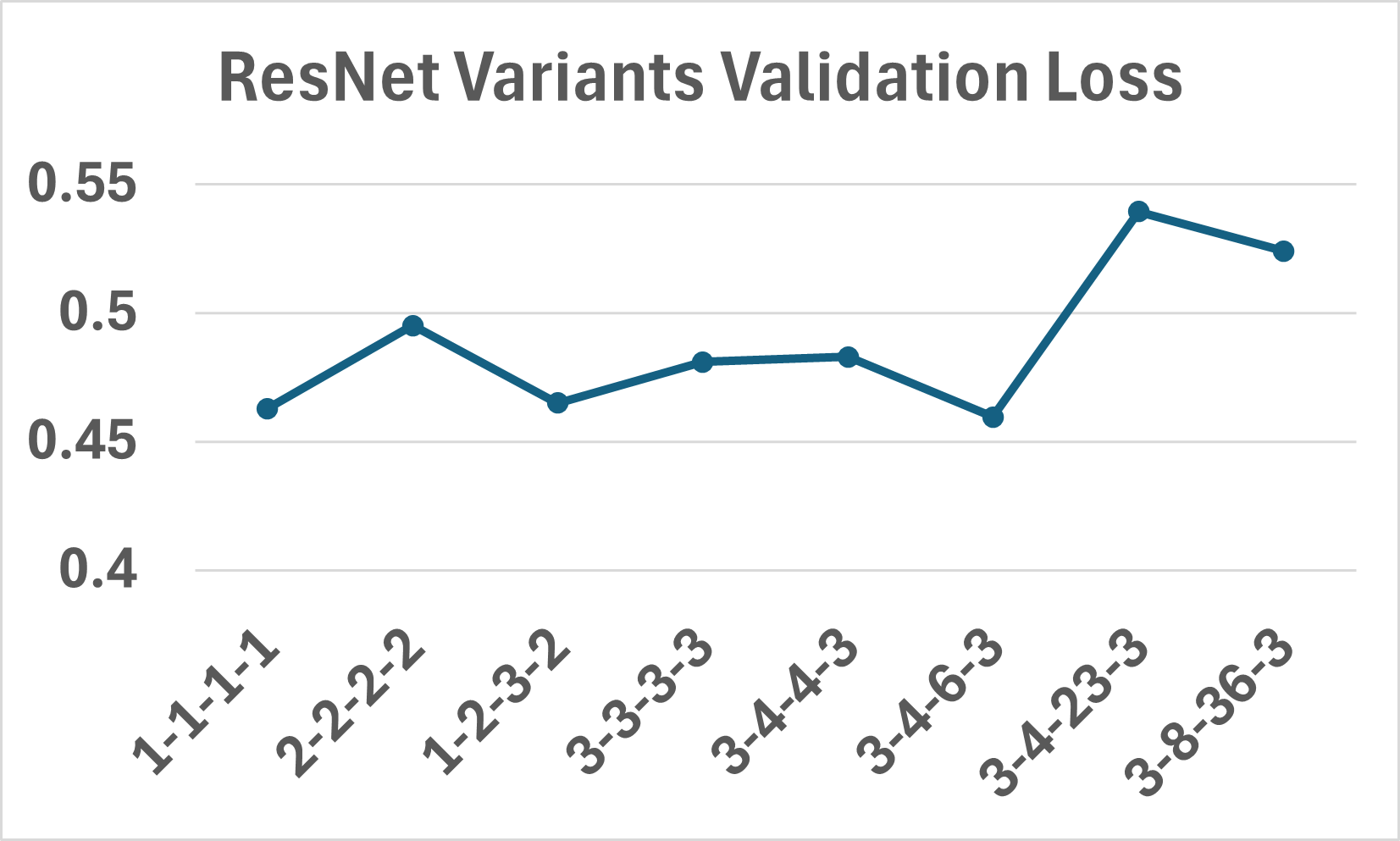}
\caption{ResNet Variants Validation Loss analysis showing the relationship between model capacity and validation performance. The validation loss does not consistently improve with increased model capacity.}
\label{fig:resnet_loss}
\end{figure}

As shown in Fig. \ref{fig:resnet_loss}, the validation loss for all ResNet variants exhibited a fluctuating pattern, with values ranging from 0.459 to 0.539. In particular, ResNet-50 achieved the lowest validation loss (0.459) among all variants, suggesting that the intermediate capacity of the model might be optimal for this specific task.
 
Training F1 scores remained consistently high (0.96-0.97) in all ResNet variants, while validation F1 scores showed minimal fluctuation (0.73-0.77), indicating significant overfitting. Even SimpleCNN with only six layers achieved comparable validation performance (F1 score 0.75) to more complex architectures, suggesting that additional model capacity does not translate to better generalization.

\begin{table*}[htb]
\scriptsize
\caption{Performance of ResNet variants on Oesophageal Adenocarcinomas Dataset.}
\label{tab:resnet}
\begin{center}
\begin{tabular}{|l|l|c|c|c|c|}
\hline
\textbf{Model} & \textbf{Blocks} & \textbf{Training Loss} & \textbf{Validation Loss} & \textbf{Training F1 score} & \textbf{Val F1 score} \\
\hline
ResNet- & [1,1,1,1] & 0.035 & 0.483 & 0.97 & 0.73 \\
ResNet-18 & [2,2,2,2] & 0.036 & 0.495 & 0.97 & 0.74 \\
ResNet- & [1,2,3,2] & 0.032 & 0.464 & 0.97 & 0.75 \\
ResNet- & [3,3,3,3] & 0.033 & 0.480 & 0.97 & 0.77 \\
ResNet- & [3,4,4,3] & 0.035 & 0.483 & 0.97 & 0.73 \\
ResNet-50 & [3,4,6,3] & 0.035 & 0.459 & 0.97 & 0.77 \\
ResNet-101 & [3,4,23,3] & 0.038 & 0.539 & 0.96 & 0.76 \\
ResNet-152 & [3,8,36,3] & 0.040 & 0.524 & 0.96 & 0.73 \\
SimpleCNN & 6 layers & 0.084 & 0.475 & 0.93 & 0.75 \\
\hline
\end{tabular}
\end{center}
\end{table*}

The gap between training and validation performance consistently indicated significant overfitting in all ResNet variants.

\subsection{Vision Transformer Analysis}
Our investigation of Vision Transformers (ViTs) revealed complex relationships between the choice of model architecture and performance. We conducted a detailed analysis of three key architectural aspects: depth, embedding size, and number of attention heads. Each of these dimensions showed distinct patterns of overfitting and performance trade-offs.

\subsubsection{Depth Impact}

\begin{figure}[htbp]
\centering
\begin{subfigure}{0.45\textwidth}
\includegraphics[width=\textwidth]{ 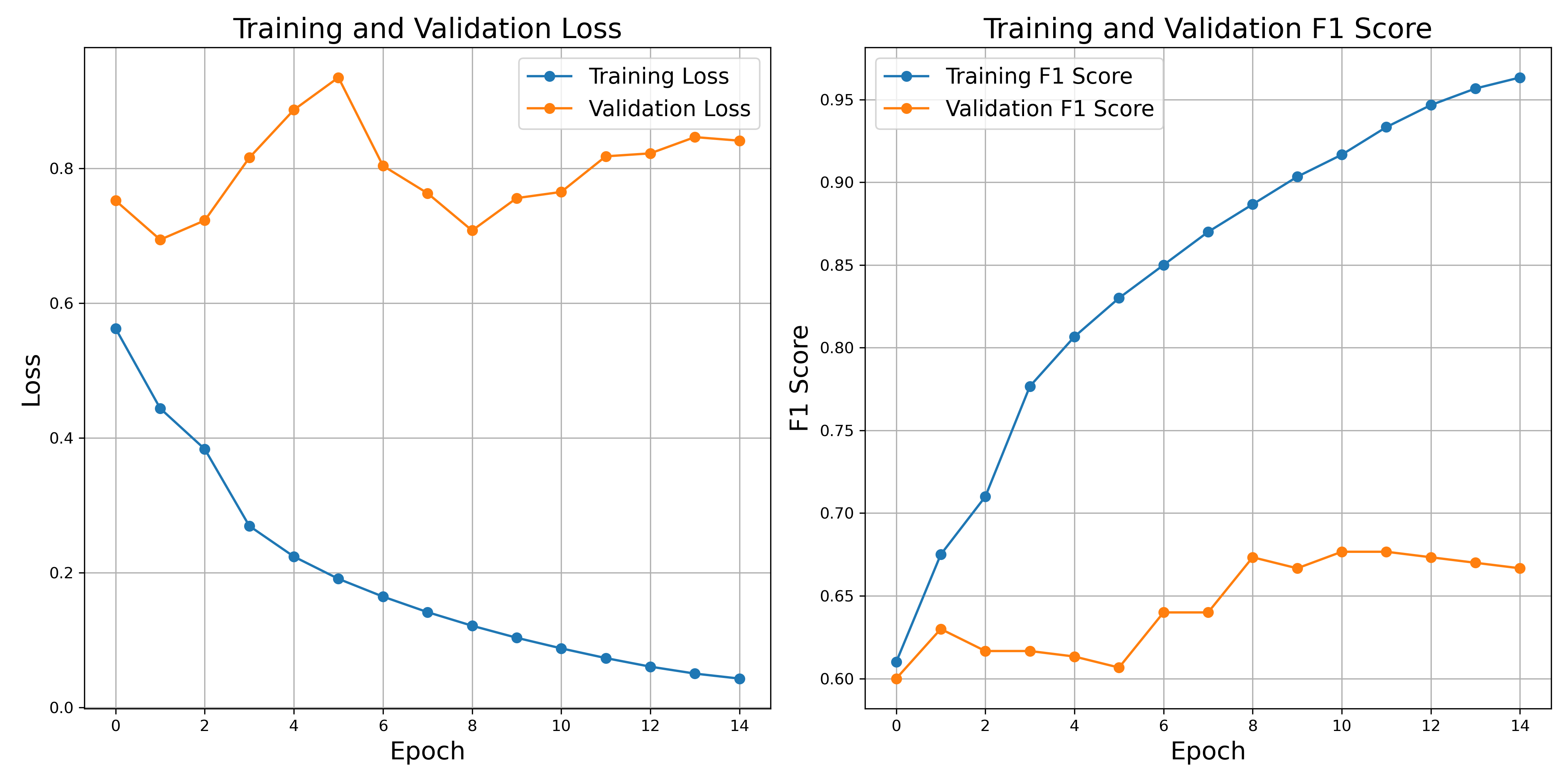}
\caption{Training curves for ViT with depth 6}
\end{subfigure}
\begin{subfigure}{0.49\textwidth}
\includegraphics[width=\textwidth]{ 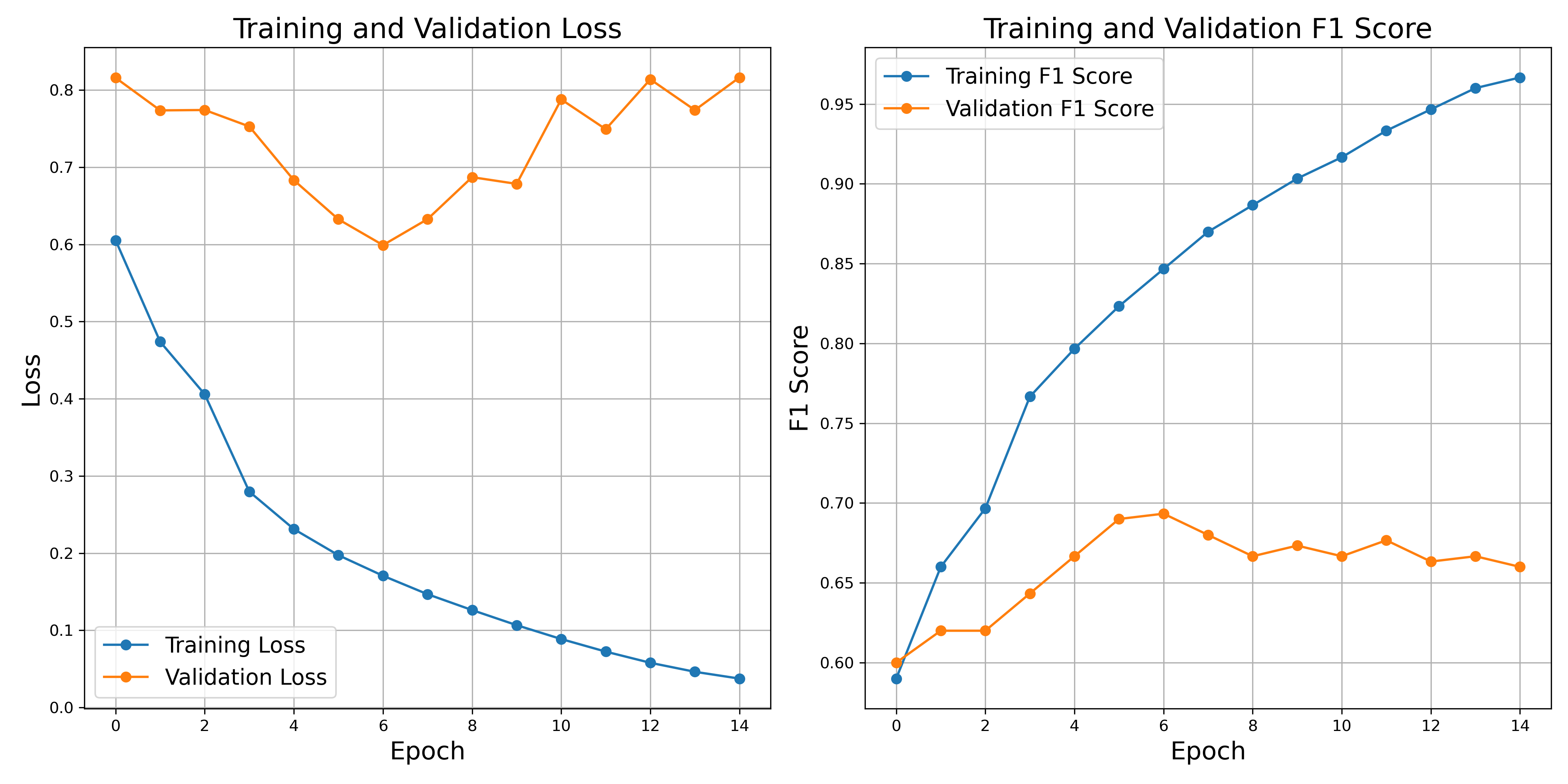}
\caption{Training curves for ViT with depth 24}
\end{subfigure}
\caption{Comparison of training dynamics between ViT models with different depths. The plots show training/validation loss and F1 scores over training epochs.}
\label{fig:vit_depth_curves}
\end{figure}

Fig. \ref{fig:vit_depth_curves} illustrates the training dynamics for ViT models with different depths. The model with a depth of 6 showed more stable training characteristics compared to the deeper variant with depth 24, which exhibited a more pronounced overfitting behavior.

Table \ref{tab:vit_depth} shows the performance metrics for various transformer depths. Increasing depth beyond 6 layers showed diminishing returns, with the 16-layer configuration achieving the best validation F1 score of 0.69.

\begin{table*}[htb]
\scriptsize
\caption{Performance of Vision Transformer with different depths.}
\label{tab:vit_depth}
\begin{center}
\begin{tabular}{|l|c|c|c|c|c|}
\hline
\textbf{Vision Transformer} & \textbf{Depth} & \textbf{Training Loss} & \textbf{Validation Loss} & \textbf{Training F1 score} & \textbf{Val F1 score} \\
\hline
ViT-Embed384-6Heads & 1 & 0.1861 & 0.8107 & 0.85 & 0.56 \\
ViT-Embed384-6Heads & 2 & 0.0813 & 0.9192 & 0.93 & 0.64 \\
ViT-Embed384-6Heads & 3 & 0.0527 & 0.909 & 0.95 & 0.65 \\
ViT-Embed384-6Heads & 4 & 0.0456 & 0.9477 & 0.96 & 0.63 \\
ViT-Embed384-6Heads & 6 & 0.0377 & 0.9004 & 0.97 & 0.66 \\
ViT-Embed384-6Heads & 8 & 0.0361 & 0.8549 & 0.97 & 0.67 \\
ViT-Embed384-6Heads & 12 & 0.0327 & 0.8765 & 0.97 & 0.66 \\
ViT-Embed384-6Heads & 16 & 0.0286 & 0.9155 & 0.98 & 0.69 \\
ViT-Embed384-6Heads & 24 & 0.0311 & 0.7972 & 0.97 & 0.66 \\
\hline
\end{tabular}
\end{center}
\end{table*}

\subsubsection{Embedding Size Effect}
The impact of the size of the embedding on the performance of the model revealed a clear pattern of diminishing returns and increased overfitting with larger embeddings. Table \ref{tab:vit_embed} presents the comprehensive results for different embedding dimensions. 

\begin{figure}[htbp]
\centering
\begin{subfigure}{0.45\textwidth}
\includegraphics[width=\textwidth]{ 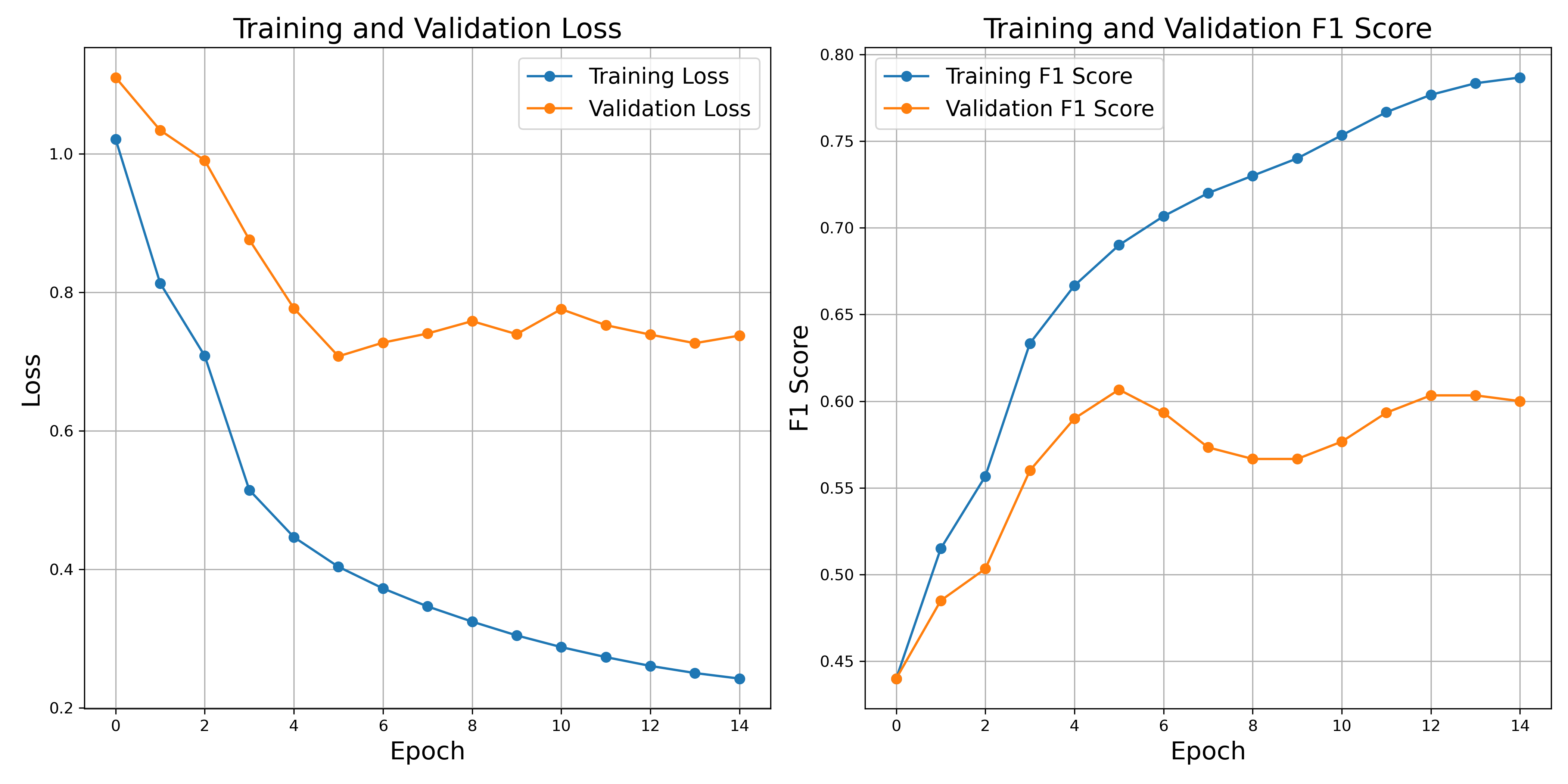}
\caption{ViT with 48-dimensional embeddings}
\end{subfigure}
\begin{subfigure}{0.49\textwidth}
\includegraphics[width=\textwidth]{ 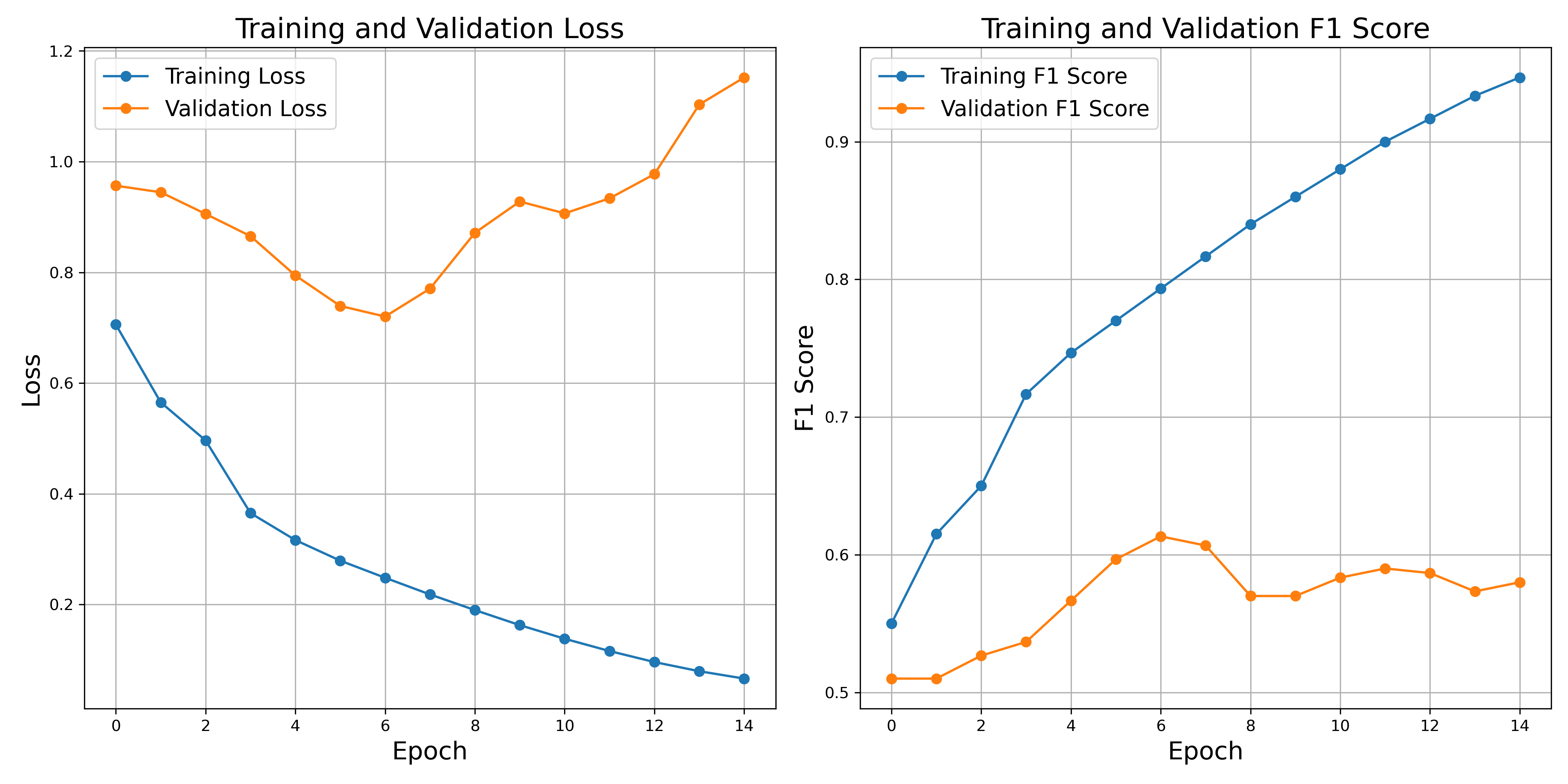}
\caption{ViT with 1536-dimensional embeddings}
\end{subfigure}
\caption{Impact of embedding dimensionality on training dynamics. The plots reveal how larger embedding sizes lead to more severe overfitting.}
\label{fig:vit_embed_compare}
\end{figure}

Fig. \ref{fig:vit_embed_compare} shows the stark contrast in training dynamics between models with small (48-dimensional) and large (1536-dimensional) embedding sizes. The smaller embedding model showed more stable validation performance, whereas the larger embedding model exhibited severe overfitting despite achieving higher training metrics.

The optimal embedding size appeared to be 384 dimensions, achieving a validation F1 score of 0.66 while maintaining reasonable training stability. Increasing the embedding size to 1536 dimensions led to a significant drop in the validation F1 score to 0.58, despite maintaining high training F1 scores of 0.96.

\begin{figure}[htbp]
\centering
\begin{subfigure}{0.45\textwidth}
\includegraphics[width=\textwidth]{ 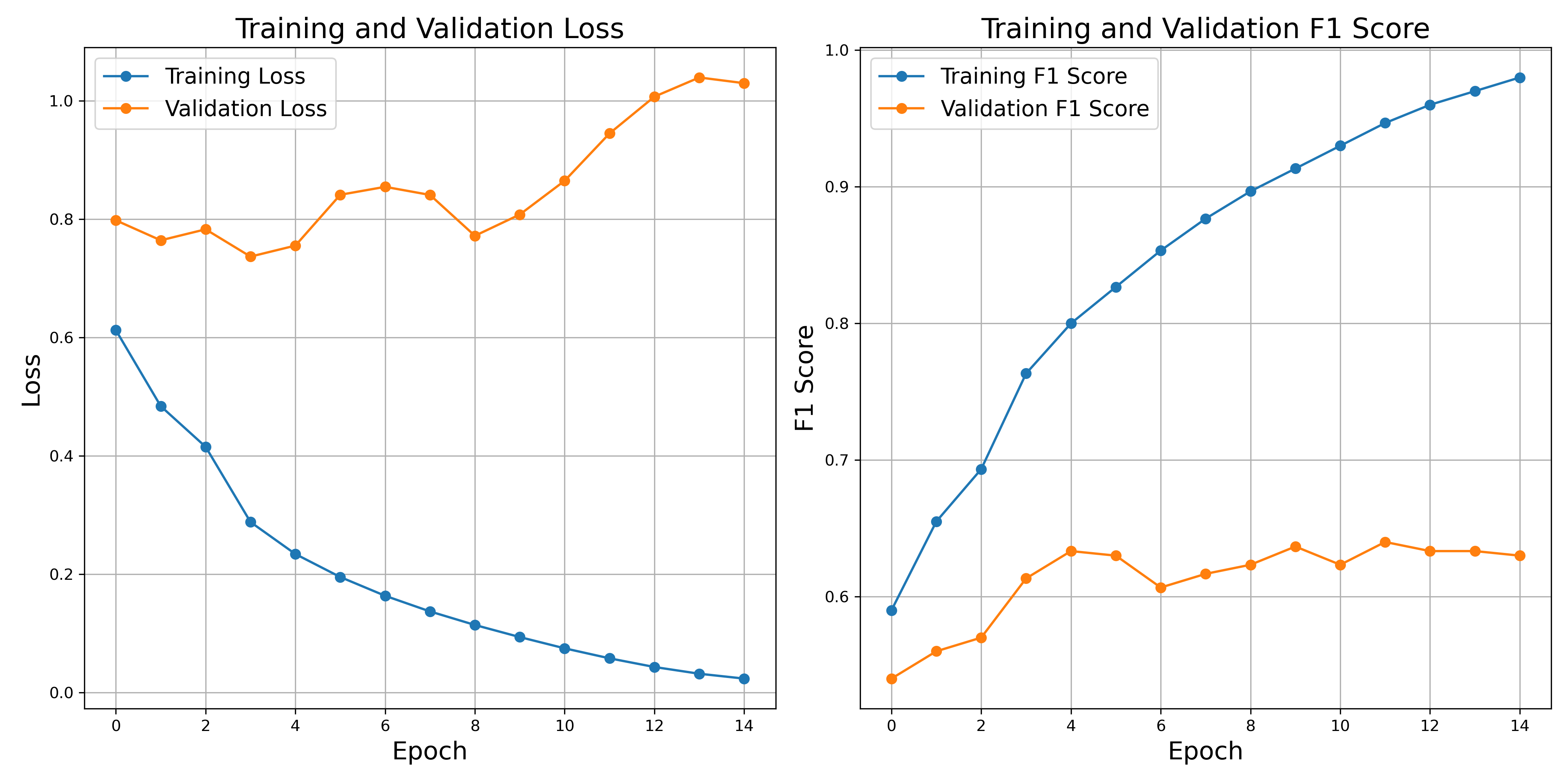}
\caption{Training curves for optimal embedding size (768)}
\end{subfigure}
\begin{subfigure}{0.49\textwidth}
\includegraphics[width=\textwidth]{ 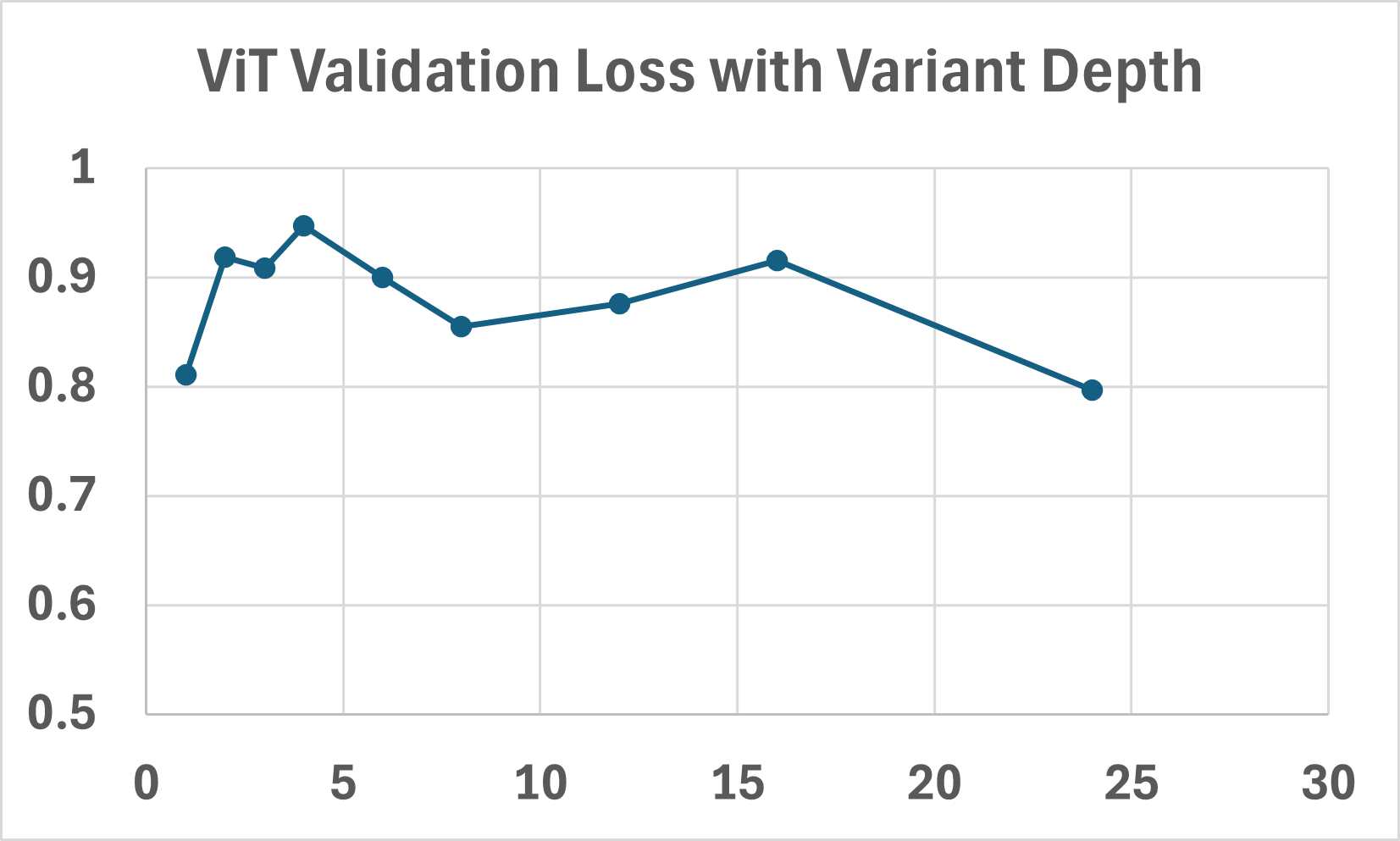}
\caption{Training curves with varying number of attention heads}
\end{subfigure}
\caption{Analysis of model behavior with optimal embedding size and attention head configurations.}
\label{fig:vit_config_compare}
\end{figure}

\begin{table*}[htb]
\scriptsize
\caption{Performance of Vision Transformer with different embedding sizes (E Sizes) and coressponding training loss (T loss), validation loss (V loss), training F1 score $F_1^T$, and validation F1 score $F_1^V$.}
\label{tab:vit_embed}
\begin{center}
\begin{tabular}{|l|c|c|c|c|c|}
\hline
\textbf{Vision Transformer} & \textbf{E size} & \textbf{T Loss} & \textbf{V Loss} & \textbf{$F_1^T$ score } & \textbf{$F_1^V$ score} \\
\hline
ViT, Depth-6, Heads-6 & 24 & 0.3383 & 0.7629 & 0.72 & 0.59 \\
ViT, Depth-6, Heads-6 & 48 & 0.2365 & 0.7127 & 0.79 & 0.61 \\
ViT, Depth-6, Heads-6 & 96 & 0.1506 & 0.8871 & 0.87 & 0.59 \\
ViT, Depth-6, Heads-6 & 192 & 0.0831 & 0.949 & 0.92 & 0.60 \\
ViT, Depth-6, Heads-6 & 384 & 0.0377 & 0.9004 & 0.97 & 0.66 \\
ViT, Depth-6, Heads-6 & 768 & 0.0235 & 0.9739 & 0.98 & 0.66 \\
ViT, Depth-6, Heads-6 & 1536 & 0.0562 & 1.1924 & 0.96 & 0.58 \\
\hline
\end{tabular}
\end{center}
\end{table*}

\section{Discussions}
Our comprehensive analysis reveals several critical insights about the application of deep learning models for histopathology. The experimental results consistently demonstrate that the common practice of adopting and scaling natural image models for histopathology tasks is suboptimal.

First, ResNet experiments (Fig. \ref{fig:resnet_loss}) showed that increasing the capacity of the model from ResNet-18 to ResNet-152 did not produce proportional improvements in validation performance. This suggests that the additional capacity of deeper networks may be unnecessary for capturing relevant features in histopathology images. In fact, as shown in Table \ref{tab:resnet}, the simplest ResNet configuration achieved a validation F1 score of 0.73, identical to that of ResNet-152, despite the latter having significantly more parameters. This pattern indicates a clear case of diminishing returns with increased model capacity.

Second, the Vision Transformer experiments revealed nuanced relationships between architectural choices and model performance. As shown in Fig. \ref{fig:vit_depth_curves}, deeper transformers (24 layers) exhibited more severe overfitting compared to moderately deep variants (6 layers). This is particularly evident in the validation loss curves, where the 24-layer model shows higher volatility and worse convergence. The training dynamics suggest that deeper architectures struggle to find stable feature representations that generalize well to unseen data.

The embedding size experiments (Fig. \ref{fig:vit_embed_compare}) demonstrated that larger embedding dimensions led to worse generalization, despite achieving higher training metrics. This phenomenon was most pronounced in the 1536-dimensional embedding model, which achieved impressive training F1 scores of 0.96 but suffered from poor validation performance (F1 score 0.58). This dramatic gap between training and validation performance suggests that larger embedding spaces enable the model to construct highly specific representations that fail to capture generalizable patterns in the data.
The attention head experiments (Fig. \ref{fig:vit_config_compare}) revealed that increasing the number of heads beyond 8 did not yield sustainable improvements in validation metrics. This finding challenges the common assumption that more attention heads automatically lead to better feature capturing capability. Instead, it suggests that excessive attention mechanisms might focus on spurious correlations in training data rather than meaningful histopathological features.

A crucial observation across all experiments is the bias-variance trade-off exhibited by different model configurations. Simpler models (fewer layers, smaller embeddings, fewer attention heads) showed higher bias but lower variance, resulting in more stable validation performance. In contrast, complex models demonstrated lower bias but significantly higher variance, leading to unstable validation metrics and poor generalization. This trade-off is particularly relevant in medical imaging applications, where reliable, consistent performance is often more valuable than occasional high-performance with low reliability. 

The dataset characteristics also play a crucial role in the observed overfitting patterns. As shown in Table \ref{tab:dataset}, our training data comes from multiple sources (SET4-CHA and SET3-TCGA) with potentially different data distributions. This heterogeneity in the training data, combined with the relatively limited sample size compared to natural image datasets, may contribute to the models' struggle to learn robust features. The validation set (SET1-UKK) might present slightly different feature distributions, exposing the models' inability to generalize across different data sources.

Furthermore, the consistent pattern of high training performance coupled with plateauing or degrading validation performance across different model architectures indicates a systemic overfitting problem. This is likely due to the fundamental mismatch between the capacity of these models and the specific characteristics of the histopathology images. Unlike natural images, which contain high semantic variability, histopathology images have more constrained features and patterns. The excess capacity of large models allows them to memorize training data rather than learn generalizable features.

These findings have important implications for both research and practical applications. They suggest that future work should focus on developing architectures specifically tailored to histopathology image analysis rather than simply adapting natural image models. Such architectures should consider the unique characteristics of histopathological data, including tissue-specific patterns, staining variations, and multi-scale features. Additionally, techniques for handling dataset bias and imbalance, such as specialized data augmentation and balanced sampling strategies, may be more crucial to improve model performance than increasing model capacity.

From a practical perspective on a small dataset, our results advocate a ``less is more'' approach in histopathology image analysis. Simpler models not only showed competitive performance, but also demonstrated more stable learning behavior and better generalization. This has significant implications for clinical applications, where model reliability and interpretability are often as important as raw performance metrics.
Looking ahead, these insights suggest several promising directions for future research, including the development of domain-specific architectural primitives, the investigation of data-efficient learning techniques, and the exploration of methods to explicitly incorporate medical domain knowledge into model design. Additionally, more research is needed to understand how different types of model capacity (depth, width, attention mechanisms) interact with specific characteristics of histopathological data.

\section{Conclusions and Future Work}
This comprehensive study illuminates the critical challenges and opportunities in applying deep learning to histopathology image analysis. Our systematic investigation of model architectures, from simple CNNs to complex Vision Transformers, reveals fundamental insights about the relationship between model capacity and performance in medical imaging tasks.

Our findings demonstrate that the current paradigm of adopting and fine-tuning natural image models for histopathology analysis is suboptimal. The experimental results consistently show that increasing model capacity for small datasets does not necessarily translate to better performance. Instead, we observed that simpler, well-designed architectures often achieve comparable or superior results while maintaining better generalization properties.

Several key conclusions emerge from our research. First, the remarkable advancements in AI for histopathology analysis are primarily data-centric rather than model-centric. The quality, size,  diversity, and representativeness of training data play a more crucial role in model performance than architectural capacity. This finding challenges the common approach of simply scaling up model size to improve performance.

Second, while deep neural networks do improve performance over traditional methods, the gains from increased model capacity are less significant than those from improved data quality and task-specific architectural design. This suggests that future research should focus more on data curation and domain-specific architectural innovations rather than simply adopting larger models from natural image processing.

Third, the limited availability of labeled histopathology data necessitates careful consideration of model architecture. Our experiments with Vision Transformers revealed more significant overfitting compared to CNNs, particularly in deeper configurations and larger embedding sizes. This indicates that architectural choices must be balanced against the constraints of available training data.

Looking ahead, several promising directions for future research emerge from our findings:
\textbf{Domain-Specific Architecture Design:} Future work should focus on developing novel architectures specifically designed for histopathology tasks. These architectures should incorporate domain knowledge about tissue structures, staining patterns, and multi-scale features characteristic of histopathology images. Potential approaches include 1) 
Development of specialized attention mechanisms that better capture local and global tissue patterns, 2) Integration of hierarchical feature learning that mirrors the multiple magnification levels used in pathology, and 3) Design of efficient architectures that maximize performance while minimizing parameter count

\textbf{Data-Efficient Learning:} Research into methods that can learn effectively from limited labeled data is crucial. This includes:

Investigation of self-supervised and semi-supervised learning techniques specific to histopathology
Development of domain-adapted pre-training strategies that better align with medical imaging characteristics
Exploration of few-shot learning approaches for rare pathological patterns

\textbf{Robust Evaluation Frameworks:} We need better ways to assess model performance in real-world clinical settings. Development of standardized benchmarks that reflect clinical requirements Creation of evaluation metrics that account for both accuracy and interoperability Implementation of testing frameworks that assess model robustness across different imaging conditions and patient populations

\textbf{Interpretability and Clinical Integration:} Future research should focus on making models more interpretable and clinically relevant.
Development of visualization techniques that align with pathologists' workflow; Integration of uncertainty quantification in model predictions; Creation of hybrid systems that combine deep learning with traditional clinical knowledge

\textbf{Multi-Modal Integration:} Future architectures should be designed to:
Incorporate multiple data modalities (e.g., genomic data, clinical history); Handle varying image resolutions and staining techniques; Support interactive refinement based on expert feedback

In conclusion, while deep learning shows great promise for histopathology image analysis, realizing this potential requires a shift away from simply adapting natural image models. Instead, success will come from the development of specialized architectures that respect the unique characteristics of histopathological data while maintaining robust performance in diverse clinical settings. This work provides a foundation for such developments and highlights the critical areas that require future research attention.

% \section*{Acknowledgments}
% To be added after review process.
\bibliographystyle{splncs04}
\bibliography{refs}
\end{document}